\documentclass[10pt,conference]{IEEEtran}
\usepackage{amsmath,amssymb}
\begin{document}
\newcommand{\E}{\mathsf{E}}
\newcommand{\Var}{\mathsf{Var}}
\newcommand{\R}{\mathbb{R}} 
\newcommand{\nn}{\nonumber}
\newcommand{\eps}{\varepsilon}
\newtheorem{theorem}{Theorem}
\renewenvironment{proof}[1][Proof]{%
\noindent\hspace{2em}{\itshape #1: }}{%
\hspace*{\fill}~\QED\par\endtrivlist\unskip}

\title{A Simple Proof of the Entropy-Power Inequality\\
via Properties of Mutual Information}
\author{
\authorblockN{Olivier Rioul}
\authorblockA{
Dept. ComElec, GET/T\'el\'ecom Paris (ENST) \\
ParisTech Institute \& CNRS LTCI\\
Paris, France\\
Email: olivier.rioul@enst.fr}
}
\maketitle

\begin{abstract}
While most useful information theoretic inequalities can be deduced from the basic properties of entropy or mutual information, Shannon's entropy power inequality (EPI) seems to be an exception: available information theoretic proofs of the EPI hinge on integral representations of differential entropy using either Fisher's information (FI) or minimum mean-square error (MMSE). In this paper, we first present a unified view of proofs via FI and MMSE, showing that they are essentially dual versions of the same proof, and then fill the gap by providing a new, simple proof of the EPI, which is solely  based on the properties of mutual information and sidesteps both FI or MMSE representations. 
\end{abstract}

\section{Introduction}

Shannon's entropy power inequality (EPI) gives a lower bound on the differential entropy of the sum of independent random variables $X,Y$ with densities:
\begin{equation}\label{epi}
\exp(2h(X+Y)) \geq \exp(2h(X)) + \exp(2h(Y))
\end{equation}
with equality if $X$ and $Y$ are Gaussian random variables. The differential entropy of the probability density function $p(x)$ of $X$ is defined as
\begin{equation}
h(X) = \E \Bigl\{ \log \frac{1}{p(X)} \Bigr\},
\end{equation}
where it is assumed throught this paper that all logarithms are natural.

The EPI finds its application in proving converses of channel or source coding theorems. It was used by Shannon as early as his 1948 paper~\cite{Shannon48} to bound the capacity of non-Gaussian additive noise channels. Recently, it was used to determine the capacity region of the Gaussian MIMO broadcast channel~\cite{WeingartenSteinbergShamai06}.
The EPI also finds application in blind source separation and deconvolution (see, e.g.,~\cite{BercherVignat00}) and is instrumental in proving a strong version of the central limit theorem with convergence in relative entropy~\cite{Barron86}.

Shannon's proof of the EPI~\cite{Shannon48} was incomplete in that he only checked that the necessary condition for a local minimum of $h(X+Y)$ is satisfied.
Available rigorous proofs of the EPI are in fact proofs of an alternative statement
\begin{equation}\label{epic}
h(\sqrt{\lambda}\,X+\sqrt{1-\lambda}\,Y)\geq \lambda h(X) + (1-\lambda) h(Y)
\end{equation}
for any $0\leq \lambda\leq 1$, which amounts to   the concavity of the entropy under the ``variance preserving'' transformation~\cite{DemboCoverThomas91}:
\begin{equation}\label{transformation}
(X,Y) \longmapsto W=\sqrt{\lambda}\,X+\sqrt{1-\lambda}\,Y.
\end{equation}
To see that \eqref{epic} is equivalent to \eqref{epi}, define $U,V$ by the relations $X=\sqrt{\lambda}\,U$, $Y=\sqrt{1-\lambda}\,V$, and rewrite \eqref{epi} as follows: 
$$
e^{2h(\sqrt{\lambda}\,U+\sqrt{1-\lambda}\,V)}\geq \lambda e^{2h(U)} + (1-\lambda)e^{2h(V)}.
$$
Taking logarithms of both sides, \eqref{epic} follows from the concavity of the logarithm. Conversely, taking exponentials, \eqref{epic} written for $U,V$ implies \eqref{epi} for $\lambda$ chosen so that $U$ and $V$ have equal entropies: $\exp 2h(U)=\exp 2h(V)$, that is, $\frac{\exp 2h(X)}{\lambda}=\frac{\exp 2h(Y)}{1-\lambda}$ or $\lambda=e^{2h(X))}/(e^{2h(X)}+e^{2h(Y)})$.

The first rigorous proof of the EPI was given by Stam~\cite{Stam59} (see also Blachman~\cite{Blachman65}). It is based on the properties of Fisher's information (FI)
\begin{equation}\label{fi}
J(X) = \E \Bigl\{ \Bigl(\frac{p'(X)}{p(X)}\Bigr)^2 \Bigr\},
\end{equation}
the link between differential entropy and FI being de Bruijn's identity~\cite[Thm.~17.7.2]{CoverThomas06}:
\begin{equation}\label{deBruijn}
\frac{d}{dt} h(X+\sqrt{t}\, Z)= \frac{1}{2}J(X+\sqrt{t}\, Z),
\end{equation}
where $Z\sim\mathcal{N}(0,1)$ is a standard Gaussian random variable, which is independent of $X$.
Recently, Verd\'u, Guo and Shamai~\cite{VerduGuo06},~\cite{GuoShamaiVerdu06a} provided an alternative proof of the EPI based on the properties of the MMSE in estimating the input $X$ to a Gaussian channel given the output $Y=\sqrt{t}\, X+Z$, where $t$ denotes the signal-to-noise ratio. This MMSE is achieved by the conditional mean estimator $\hat{X}(Y)=\E(X|Y)$ and is given by the conditional variance
\begin{equation}
\Var(X|Y) = \E \Bigl\{ (X-\E\{X|Y\})^2\Bigr\},
\end{equation}
where the expectation is taken over the joint distribution of the random variables $X$ and $Y$.
The connection between input-output mutual information $I(X;Y)=h(Y)-h(Z)$ and MMSE is made by the following identity derived in~\cite{GuoShamaiVerdu05}:
\begin{equation}\label{GuoVerduShamai}
\frac{d}{dt} I(X;\sqrt{t}\,  X + Z) = \frac{1}{2} \Var(X|\sqrt{t}\,  X + Z).
\end{equation}
This identity turns out to be equivalent to de Bruijn's identity~\eqref{deBruijn}. It has been claimed~\cite{GuoShamaiVerdu06a} that using the alternative MMSE representation in place of FI representation is more insightful and convenient for proving the EPI.

In this paper, we show that it is possible to avoid both MMSE and FI representations and use only basic properties of mutual information. The new proof of the EPI presented in this paper is based on a convexity inequality for mutual information under the variance preserving transformation~\eqref{transformation}:
\begin{theorem}\label{R}
If $X$ and $Y$ are independent random variables, and if $Z$ is Gaussian independent of $X,Y$, then
\begin{align}\label{rioul}
I(\sqrt{\lambda}\,X&+\sqrt{1-\lambda}\,Y+\sqrt{t}\,Z;Z)\nn\\
&\leq \lambda I(X+\sqrt{t}\,Z;Z)+(1-\lambda) I(Y+\sqrt{t}\,Z;Z)
\end{align}
for all $0\leq \lambda \leq 1$ and $t\geq 0$.
\end{theorem}
Apart from its intrinsic interest, we show that inequality~\eqref{rioul} reduces to the EPI by letting $t\to\infty$.

Before turning to the proof of Theorem~\ref{R}, we make the connection between earlier proofs of the EPI via FI and via MMSE by focusing on the essential ingredients common to the proofs. This will give an idea of the level of difficulty that is required to understand the conventional approaches, while also serving as a guide to understand the new proof which uses similar ingredients, but is comparatively simpler and shorter.

The remainder of the paper is organized as follows. Section~\ref{FIMMSE} gives a direct proof of a simple relation between FI and MMSE, interprets~\eqref{deBruijn} and~\eqref{GuoVerduShamai} as dual consequences of a generalized identity, and explores the relationship between the two previous proofs of the EPI via FI and via MMSE. 
It is shown that these are essentially dual versions of the same proof; they follow the same lines of thought and each step has an equivalent formulation, and a similar interpretation, in terms of FI and MMSE. 
Section~\ref{EPII} then proves Theorem~\ref{R} and the EPI using two basic ingredients common to earlier approaches, namely 1) a ``data processing'' argument applied to~\eqref{transformation}; 2) a Gaussian perturbation method. The reader may wish to skip to this section first, which does not use the results presented earlier. The new approach has the advantage of being very simple in that it relies only on the basic properties of mutual information.

\section{Proofs of the EPI via FI and MMSE revisited}\label{FIMMSE}

The central link between FI and MMSE takes the form of a simple relation which shows that they are complementary quantities in the case of a standard Gaussian perturbation~$Z$ independent of $X$: 
\begin{equation}\label{fimmse}
J(X+Z)+\Var(X|X+Z)=1
\end{equation}
This identity  was mentioned in~\cite{GuoShamaiVerdu05} to show that~\eqref{deBruijn} and~\eqref{GuoVerduShamai} are equivalent. We first provide a direct proof of this relation, and then use it to unify and simplify existing proofs of the EPI via FI and via MMSE.
In particular, two essential ingredients, namely, Fisher's information inequality~\cite{Blachman65},~\cite{Zamir98}, and a related inequality for MMSE~\cite{VerduGuo06},~\cite{GuoShamaiVerdu06a}, will be shown to be equivalent from~\eqref{fimmse}.

\subsection{A new proof of~\eqref{fimmse}}
Fisher's information~\eqref{fi} can be written in the form
\begin{equation}
J(X) = \E\{ S^2(X)\} = \Var\{S(X)\}
\end{equation}
where $S(X)=p'(X)/p(X)$ is a zero-mean random variable.
The following conditional mean representation is due to Blachman~\cite{Blachman65}:
\begin{equation}\label{blach}
S(X+Z)=\E\{S(Z)|X+Z\}.
\end{equation}
By the ``law of total variance'', this gives
\begin{align}
J(X+Z) &= \Var\{S(X+Z)\}\nn\\
&=  \Var\{\E\{S(Z)|X+Z\}\}\nn\\
&= \Var\{S(Z)\} - \Var\{S(Z)|X+Z\}\nn\\
&= J(Z) - \Var\{S(Z)|X+Z\} \label{a}
\end{align}
We now use the fact that $Z$ is standard Gaussian. It is easily seen by direct calculation that $S(Z)=-Z$ and $J(Z)=1$, and, therefore, $J(X+Z) = 1 - \Var\{Z|X+Z\}$.
Since $Z-\E(Z|X+Z)=\E(X|X+Z)-X$ we have $\Var\{Z|X+Z\}=\Var\{X|X+Z\}$, thereby showing~\eqref{fimmse}.\endproof
Note that when $Z$ is Gaussian but not standard Gaussian, we have $S(Z)=-Z/\Var(Z)$ and $J(Z)=1/\Var(Z)$, and~\eqref{fimmse} generalizes to
\begin{equation}\label{fimmseg}
\Var(Z)J(X+Z)+J(Z)\Var(X|X+Z)=1.
\end{equation}
Another proof, which is based on a data processing argument and avoids Blachman's representation, is given in~\cite{Rioul07}.

\subsection{Intepretation}

Equation~\eqref{fimmse} provides a new estimation theoretic interpretation of Fisher's information of a noisy version $X'=X+Z$ of $X$. It is just the complementary quantity to the MMSE that results from estimating $X$ from $X'$. The estimation is all the more better as the MMSE is lower, that is, as $X'$ provides higher FI. Thus Fisher's information is a measure of least squares estimation's efficiency, when estimation is made in additive Gaussian noise.

To illustrate, consider the special case of a Gaussian random variable $X$. Then the best estimator is the linear regression estimator, with MMSE equal to $\Var(X| X') = (1-\rho^2) \Var(X)$ where $\rho=\sqrt{\Var(X)/\Var(X')}$ is the correlation factor between $X$ and $X'$:
\begin{equation}
\Var(X| X') = \frac{\Var(X)}{\Var(X)+1}.
\end{equation}
Meanwhile, $J(X')$ is simply the reciprocal of the variance of~$X'$:
\begin{equation}
J(X') = \frac{1}{\Var(X)+1}.
\end{equation}
Both quantities sum to one, in accordance with~\eqref{fimmse}.
In the case of non-Gaussian noise, we have the more general identity~\eqref{a} which also links Fisher information and conditional variance, albeit in a more complicated form. 

\subsection{Dual versions of de Bruijn's Identity}

De Bruijn's identity can be stated in the form~\cite{DemboCoverThomas91}
\begin{equation}\label{deBruijn0}
\frac{d}{dt} h(X+\sqrt{t}\,  Z)\Bigl|_{t=0}= \frac{1}{2}J(X)\Var(Z).
\end{equation}
The conventional technical proof of~\eqref{deBruijn0} is obtained by integrating by parts using a diffusion equation satisfied by the Gaussian distribution (see e.g.,~\cite{Blachman65} and~\cite[Thm.~17.7.2]{CoverThomas06}). A simpler proof of a more general result is included in~\cite{Rioul07}. From~\eqref{deBruijn0}, we deduce the following.
\begin{theorem}[de Bruijn's identity]
For any two random independent random variables $X$ and $Z$,
\begin{align}\label{dB1}
\frac{d}{dt} h(X+\sqrt{t}\, Z) &= \frac{1}{2}\,J(X+\sqrt{t}\, Z)\,\Var(Z)\\
\intertext{if $Z$ is Gaussian, and}\label{dB2}
\frac{d}{dt} h(X+\sqrt{t}\, Z) &= \frac{1}{2}\,J(X)\;\Var(Z|X+\sqrt{t}\, Z)
\end{align}
if $X$ is Gaussian.
\end{theorem}

This theorem is essentially contained in~\cite{GuoShamaiVerdu05}. 
In fact, noting that $I(X;\sqrt{t}\, X+Z)=h(\sqrt{t}\,X+Z)-h(Z)$, it is easily seen that~\eqref{dB2}, with $X$ and $Z$ interchanged, is the  identity~\eqref{GuoVerduShamai} of Guo, Verd\'u and Shamai. Written in the form~\eqref{dB2} it is clear that this is a dual version of the conventional de Bruijn's identity~\eqref{dB1}.
Note that both identities reduce to~\eqref{deBruijn0} for $t=0$. For completeness we include a simple proof for $t>0$.

\begin{proof}
Equation~\eqref{dB1} easily follows from~\eqref{deBruijn0} using the stability property of Gaussian distributions under convolution: substitute $X+\sqrt{t'}\,Z'$ for $X$ in~\eqref{deBruijn0}, where $Z$ and $Z'$ are taken to be iid Gaussian random variables, and use the fact that $\sqrt{t}\,Z+\sqrt{t'}\,Z'$ and $\sqrt{t+t'}\,Z$ are identically distributed.

To prove~\eqref{dB2} we use the complementary relation~\eqref{fimmseg} in the form
\begin{equation}\label{b}
\Var(X)J(X+\sqrt{t}\,Z)+tJ(X)\Var(Z|X+\sqrt{t}\,Z)=1
\end{equation}
where $X$ is Gaussian. Let $u=1/t$. By~\eqref{dB1} (with $X$ and $Z$ interchanged), we have
\begin{align}
\frac{d}{dt} h(X+\sqrt{t}\, Z) &= \frac{d}{dt} \Bigl\{ h(\sqrt{u}\,X+ Z) + \frac{1}{2}\log t\Bigr\}\nn\\
&= -\frac{1}{t^2}\frac{ d}{du}h(\sqrt{u}\,X+ Z) + \frac{1}{2t} \nn\\
&= -\frac{1}{2t^2} \Var(X) J(\sqrt{u}\,X+Z) + \frac{1}{2t}\nn\\
&= -\frac{1}{2t} \Var(X) J(X+\sqrt{t}\,Z) + \frac{1}{2t}.\nn
\end{align}
which combined with~\eqref{b} proves~\eqref{dB2}.
\end{proof}

\subsection{Equivalent integral representations of differential entropy}

Consider any random variable with density and finite variance $\sigma^2=\Var(X)$.
Its non-Gaussianness is defined as the divergence with respect to a Gaussian random variable $X_G$ with identical second centered moments, and is given by
\begin{equation}\label{h}
D_h(X) = h(X_G)-h(X)
\end{equation}
where $h(X_G)=\frac{1}{2}\log \bigl( 2\pi e \sigma^2\bigr)$. Let $Z$ be standard Gaussian, independent of $X$. From~\eqref{dB1}, we obtain
\begin{equation}\label{c}
\frac{d}{dt} D_h(X+\sqrt{t}\,Z) = -\frac{1}{2} D_J(X+\sqrt{t}\,Z),
\end{equation}
where
\begin{equation}\label{DJ}
D_J(X)=J(X)-J(X_G).
\end{equation}
Here $J(X_G)=1/\sigma^2$ and~\eqref{DJ} is nonnegative by the Cram\'er-Rao inequality.
Now for $t=0$, $D_h(X+\sqrt{t}\,Z)=D_h(X)$, and since non-Gaussianness is scale invariant, $D_h(X+\sqrt{t}\,Z)=D_h(Z+X/\sqrt{t})\to D_h(Z)=0$ as $t\to+\infty$. 
Therefore, integrating~\eqref{c} from $t=0$ to $+\infty$ we obtain a FI integral representation for differential entropy~\cite{Barron86}
\begin{equation}\label{e}
D_h(X) = \frac{1}{2}\int_0^{\infty} D_J(X+\sqrt{t}\, Z) \,dt
\end{equation}
or
\begin{equation}\label{fir}
h(X) = \frac{1}{2}\log \bigl( 2\pi e \sigma^2\bigr) - \frac{1}{2}\int_0^\infty\!\!\!\!J(X+\sqrt{t}\,Z)-\frac{1}{\sigma^2+t} \, dt.
\end{equation}
Similarly, from~\eqref{dB2} with $X$ and $Z$ interchanged, we obtain a dual identity:
\begin{equation}\label{d}
\frac{d}{dt} D_h(\sqrt{t}\,X+Z) = \frac{1}{2} D_V(X|\sqrt{t}\,X+Z),
\end{equation}
where for $Y=\sqrt{t}\,X+Z$ and $Y_G=\sqrt{t}\,X_G+Z$,
\begin{equation}
D_V(X|Y)=\Var(X_G|Y_G)-\Var(X|Y).
\end{equation}
Again this quantity is nonnegative, because $\Var(X_G|Y_G)=\sigma^2/(t\sigma^2+1)$ is the MMSE achievable by a linear estimator, which is suboptimal for non-Gaussian~$X$. Now for $t=0$, $D_h(\sqrt{t}\,X+Z)=D_h(Z)$ vanishes, and since non-Gaussianness is scale invariant, $D_h(\sqrt{t}\,X+Z)=D_h(X+Z/\sqrt{t})\to D_h(X)$ as $t\to+\infty$. Therefore, integrating~\eqref{d} from $t=0$ to $+\infty$ we readily obtain the MMSE integral representation for differential entropy~\cite{GuoShamaiVerdu05}:
\begin{equation}\label{f}
D_h(X) = \frac{1}{2}\int_0^{\infty} D_V(X|\sqrt{t}\,X+Z) \,dt
\end{equation}
or
\begin{equation}\label{mmser}
h(X) = \frac{1}{2}\log \bigl( 2\pi e \sigma^2\bigr) - \frac{1}{2}\int_0^\infty\!\!\!\!
\frac{\sigma^2}{1+t\sigma^2} -\Var(X|\sqrt{t}\,X+Z) \, dt.
\end{equation}
This MMSE representation was first derived in~\cite{GuoShamaiVerdu05} and used in~\cite{VerduGuo06},~\cite{GuoShamaiVerdu06a} to prove the EPI. The FI representation~\eqref{fir} can be used similarly, yielding essentially Stam's proof of the EPI~\cite{Stam59},~\cite{Blachman65}. These proofs are sketched below. 

Note that the equivalence between FI and MMSE representations~\eqref{e},~\eqref{f} is immediate by the complementary relation~\eqref{fimmse}, which can be simply written as
\begin{equation}\label{g}
D_J(X+Z)=D_V(X|X+Z).
\end{equation}
In fact, it is easy to see that~\eqref{e} and~\eqref{f} prove each other by~\eqref{g} after making a change of variable $u=1/t$.
Both sides of~\eqref{g} are of course simultaneously nonnegative and measure a ``non-Gaussianness'' of $X$ when estimated in additive Gaussian noise. Interestingly, these FI and MMSE non-Gaussianities coincide.

It is also useful to note that in the above derivations, the Gaussian random variable $X_G$ may very well be chosen such that $\sigma^2=\Var(X_G)$ is not equal to $\Var(X)$. Formulas~\eqref{h}--\eqref{g} still hold, even though quantities such as $D_h$, $D_J$, and $D_V$ may take negative values. In particular, the right-hand sides of~\eqref{fir} and~\eqref{mmser} do not depend on the particular choice of $\sigma$.

\subsection{Simplified proofs of the EPI via FI and via MMSE}

For Gaussian random variables $X_G$, $Y_G$ with the same variance, the EPI in the form~\eqref{epic} holds trivially with equality. Therefore, to prove the EPI, it is sufficient to show the following convexity property:
\begin{equation}\label{epic2}
D_h(W)\leq \lambda D_h(X) + (1-\lambda) D_h(Y)
\end{equation}
where $D_h(\cdot)$ is defined by~\eqref{h} and $W$ is defined as in~\eqref{transformation}. By the last remark of the preceding subsection, the FI and MMSE representations~\eqref{e},~\eqref{f} hold. Therefore, to prove the EPI, it is sufficient to show that either one of the following inequalities holds.
\begin{align*}
D_J(W&+\sqrt{t}\,Z)\\
& \leq\lambda D_J(X+\sqrt{t}\,Z) + (1-\lambda) D_J(Y+\sqrt{t}\,Z)\\
D_V(W&|\sqrt{t}\,W+Z)\\
& \leq\lambda D_V(X|\sqrt{t}\,X+Z) + (1-\lambda) D_V(Y|\sqrt{t}\,Y+Z).
\end{align*}
These in turn can be written as
\begin{align}
J(W&+\sqrt{t}\,Z) \nn\\\label{j}
& \leq\lambda J(X+\sqrt{t}\,Z) + (1-\lambda) J(Y+\sqrt{t}\,Z)\\
\Var(W&|\sqrt{t}\,W+Z)\nn\\ \label{k}
& \geq\lambda \Var(X|\sqrt{t}\,X+Z) + (1-\lambda) \Var(Y|\sqrt{t}\,Y+Z),
\end{align}
because these inequalities hold trivially with equality for $X_G$ and $Y_G$.

Inequality~\eqref{k} is easily proved in the form
\begin{equation}\label{l}
\Var(W|\sqrt{t}\,W+Z) \geq \Var(W|\sqrt{t}\,X+Z',\sqrt{t}\,Y+Z'' )
\end{equation}
where $Z'$ and $Z''$ are standard Gaussian random variables, independent of $X,Y$ and of each other, and $Z=\sqrt{\lambda}\,Z'+\sqrt{1-\lambda}\,Z''$.
This has a simple interpretation~\cite{VerduGuo06},~\cite{GuoShamaiVerdu06a}: it is better to estimate the sum of independent random variables from individual noisy measurements than from the sum of these measurements.

The dual inequality~\eqref{j} can also be proved in the form
\begin{equation}
J(W) \leq\lambda J(X) + (1-\lambda) J(Y)\\
\end{equation}
where we have substituted $X,Y$ for $X+\sqrt{t}\,Z', Y+\sqrt{t}\,Z''$. This inequality is known as the Fisher's information inequality~\cite{DemboCoverThomas91}, and an equivalent formulation is
\begin{equation}\label{m}
\frac{1}{J(X+Y)} \geq \frac{1}{J(X)} + \frac{1}{J(Y)}.
\end{equation}
Blachman~\cite{Blachman65} gave a short proof using representation~\eqref{blach} and Zamir~\cite{Zamir98} gave an insightful proof using a data processing argument. Again~\eqref{m} has a simple interpretation, which is very similar to that of~\eqref{l}. Here $1/J(X)$ is the Cram\'er-Rao lower bound (CRB) of the mean-squared error of the unbiaised estimator $X$ for a translation parameter, and~\eqref{m} states that in terms of the CRB it is better to estimate the translation parameter corresponding to the sum of independent random variables $X+Y$ from individual measurements than from the sum of these measurements.

At any rate, the equivalence between~\eqref{j} and~\eqref{k} is immediate by the complementary relation~\eqref{fimmse} or its generalized version~\eqref{fimmseg}, as can be easily seen. Either one of~\eqref{j},~\eqref{k} gives a proof of the EPI.\endproof

The above derivations illuminate intimate connections between both proofs of the EPI, via FI and via MMSE. They do not only follow the same lines of argumentation, but are also shown to be dual in the sense that through~\eqref{fimmse},  each step in these proofs has an equivalent formulation and similar interpretation in terms of FI or MMSE.

\section{A new proof of the EPI}\label{EPII}

For convenience in the following proof, define $W$ by~\eqref{transformation} and let $\alpha=\sqrt{t\lambda}$ and $\beta=\sqrt{t(1-\lambda)}$. 

\subsection{Proof of Theorem~\ref{R}}

Similarly as in the conventional proofs of the EPI, we use the fact that the linear combination $\sqrt{\lambda} (X+\alpha Z)+\sqrt{1-\lambda} (Y+\beta Z) = W + \sqrt{t}\,Z$ cannot bring more information than the individual variables $X+\alpha Z$ and $Y+\beta Z$ together. Thus, by the data processing inequality for mutual information,
\begin{equation}\label{dpimi}
I(W+\sqrt{t}\,Z;Z)\leq I(X+ \alpha Z,Y+ \beta Z;Z).
\end{equation}
Let $U=X+\alpha Z$, $V=Y+\beta Z$ and  develop using the chain rule for mutual information:
\begin{align*}
I(U,V;Z) &= I(U;Z)+I(V;Z|U)\\
&\leq I(U;Z) + I(V;Z|U) + I(U;V)\\
&= I(U;Z) + I(V;U,Z)\\
&=I(U;Z) + I(V;Z) + I(U;V|Z).
\end{align*}
Since $X$ and $Y$ are independent, $U$ and $V$ are conditionally independent given $Z$, and therefore, $I(U;V|Z)=0$. Thus, we obtain the inequality
\begin{equation}\label{r}
I(W+\sqrt{t}\,Z;Z)\leq I(X+ \alpha Z;Z)+I(Y+ \beta Z;Z)
\end{equation}
Assume for the moment that $I(X+\alpha Z;Z)$ admits a second-order Taylor expansion about $\alpha=0$ as $t\to 0$. Since $I(X+ \alpha Z;Z)$ vanishes for $\alpha=0$, and since mutual information is nonnegative, we may write
\begin{equation*}
I(X+\alpha Z; Z) = \lambda I(X+\sqrt{t}\,Z;Z) + o(\alpha^2)
\end{equation*}
where $o(\alpha^2)=o(t)$ and $\frac{o(t)}{t}$ tends to zero as $t\to 0$.
Similarly $I(Y+ \beta Z;Z)=(1-\lambda)I(Y+\sqrt{t}\,Z;Z) + o(t)$. It follows that in the vicinity of $t=0$, 
\begin{multline}\label{ot}
I(W+\sqrt{t}\,Z;Z) \leq \lambda I(X+\sqrt{t}\,Z;Z)\\+(1-\lambda)I(Y+\sqrt{t}\,Z;Z) + o(t).
\end{multline}
We now remove the $o(t)$ term by using the assumption that $Z$ is Gaussian. Consider the variables $X'=X+\sqrt{t'}\,Z'_1$, $Y'=Y+\sqrt{t'}\,Z'_2$, where $Z'_1,Z'_2$ are identically distributed as $Z$ but independent of all other random variables. This Gaussian perturbation ensures that densities are smooth, so that $I(X'+\alpha Z;Z)$ and $I(Y'+\beta Z;Z)$ both admit a second-order Taylor expansion about $t=0$. We may, therefore, apply~\eqref{ot} to $X'$ and $Y'$, which gives 
\begin{multline}\label{ot2}
I(W+\sqrt{t'}\,Z'+\sqrt{t}\,Z;Z) \leq \lambda I(X+\sqrt{t'}\,Z'_1+\sqrt{t}\,Z;Z)\\+(1-\lambda)I(Y+\sqrt{t'}\,Z'_2+\sqrt{t}\,Z;Z)+ o(t)
\end{multline}
where $Z'=\sqrt{\lambda}\, Z'_1+\sqrt{1-\lambda}\, Z'_2$ is identically distributed as $Z$.
Applying the obvious identity $I(X+Z'+Z;Z)=I(X+Z'+Z;Z'+Z)-I(X+Z';Z')$ and using the stability property of the Gaussian distribution under convolution, \eqref{ot2} boils down to 
$$
f(t'+t)\leq f(t')+o(t)
$$
where we have noted $f(t)=I(W+\sqrt{t}\,Z;Z)-\lambda I(X+\sqrt{t}\,Z;Z)-(1-\lambda) I(Y+\sqrt{t}\,Z;Z)$. It follows that $f(t)$ is non increasing in $t$, and since it clearly vanishes for $t=0$, it always assumes non positive values for all $t\geq 0$. This completes the proof of~\eqref{rioul}.
\endproof

Interestingly, this proof uses two basic ingredients common to earlier proofs presented in section~\ref{FIMMSE}: 1) the fact that two variables together bring more information than their sum, which is here expressed as a data processing inequality for mutual information; 2) a Gaussian perturbation argument using an auxiliary variable $Z$.

In fact, Theorem~\ref{R} could also be proved using either one of the integral representations~\eqref{fir},~\eqref{mmser}, which are equivalent by virtue of~\eqref{fimmse} and obtained through de Bruijn's identity as explained in section~\ref{FIMMSE}. The originality in the present proof is that it does neither require de Bruijn's identity nor the notions of FI or MMSE.

\subsection{The discrete case}

The above proof of Theorem~\ref{R} does not require that $X$ or $Y$ be random variables with densities. Therefore, Theorem~\ref{R} also holds for discrete (finitely or countably) real valued random variables. Verd\'u and Guo~\cite{VerduGuo06} proved that the EPI, in the form~\eqref{epic}, also holds in this case, where differential entropies are replaced by entropies. 
We call attention that this is in fact a trivial consequence of the stronger inequality
\begin{equation}
H(\sqrt{\lambda}\,X+\sqrt{1-\lambda}\,Y) \geq \max\bigl(H(X),H(Y)\bigr)
\end{equation}
for any two independent discrete random variables $X$ and $Y$.
This inequality is easily obtained by noting that 
\begin{equation}
H(W) \geq H(W|Y)=H(X|Y)=H(X)
\end{equation}
and similarly for $H(Y)$.

\subsection{Proof of the EPI}

We now show that the EPI for differential entropies, in the form~\eqref{epic}, follows from Theorem~\ref{R}. By the identity $I(X+Z;Z)+h(X)=I(X;X+Z)+h(Z)$, inequality~\eqref{rioul} can be written in the form
\begin{multline}\label{vg}
h(W)- \lambda h(X) - (1-\lambda) h(Y) \geq 
I(W;W+\sqrt{t}\,Z)\\- \lambda I(X;X+\sqrt{t}\,Z) - (1-\lambda) I(Y;Y+\sqrt{t}\,Z).
\end{multline}
We now let $t\to\infty$ in the right-hand side of this inequality.
Let $\eps=1/\sqrt{t}$ and $X_G$ be a Gaussian random variable independent of $Z$, with identical second moments as $X$. Then $I(X;X+\sqrt{t}\,Z)=I(X;\eps X+Z)=h(\eps X+Z)-h(Z)\leq h(\eps X_G+ Z)-h(Z)=\frac{1}{2}\log (1+{\sigma^2_X}/{t\sigma^2_Z})$, which tends to zero as $t\to\infty$. This holds similarly for the other terms in the right-hand side of~\eqref{vg}. Therefore, the EPI~\eqref{epic} follows.
\endproof

In light of this proof, we see that theorem~\ref{R}, which contains the EPI as the special case where $\sigma^2_Z\to\infty$, merely states that the difference $h(W+Z)-\lambda h(X+Z)-(1-\lambda) h(Y+Z)$ between both sides of the EPI~\eqref{epic} decreases as independent Gaussian noise $Z$ is added. This holds in accordance
with the fact that this difference is zero for Gaussian random variables with identical variances.

One may wonder if mutual informations in the form $I(X;\sqrt{t}\,X+Z)$ rather than $I(X+\sqrt{t}\,Z;Z)$ could be used in the above derivation of Theorem~\ref{R} and the EPI, in a similar manner as Verd\'u and Guo's proof uses~\eqref{mmser} rather than~\eqref{fir}.
In fact, this would amount to proving~\eqref{vg}, whose natural derivation using the data processing inequality for mutual information is through~\eqref{dpimi}. 

The same proof we used above can be employed verbatim to prove the EPI for random vectors. Generalizations to various extended versions of EPI are provided in a follow-up to this work~\cite{Rioul07}.

\end{document}